# Enhanced Optical Chirality through Locally Excited Surface Plasmon Polaritons

M.H. Alizadeh[‡,¶] and Björn M. Reinhard[†,¶]

[‡]Department of Physics, [†]Department of Chemistry, and [¶]The Photonics Center, Boston University, Boston, MA 02215, United States

*Abstract:*

*Plasmonic nanostructures provide unique opportunities to improve the detection limits of chiroptical spectroscopies by enhancing chiral light-matter interactions. The most significant of such interaction occur in ultraviolet (UV) range of the electromagnetic spectrum that remains challenging to access by conventional localized plasmon resonance based sensors. Although Surface Plasmon Polaritons (SPPs) on noble metal films can sustain resonances in the desired spectral range, their transverse magnetic nature has been an obstacle for enhancing chiroptical effects. Here we demonstrate, both analytically and numerically, that SPPs excited by near-field sources can exhibit rich and non-trivial chiral characteristics. In particular, we show that the excitation of SPPs by a chiral source not only results in a locally enhanced optical chirality but also achieves manifold enhancement of net optical chirality. Our finding that SPPs facilitate a plasmonic enhancement of optical chirality in the UV part of the spectrum is of great interest in chiral bio-sensing.*

Surface plasmon polaritons, SPPs, are surface electromagnetic excitations confined to and propagating along metal-dielectric interfaces[1, 2]. Upon coupling to photons, a nonzero local charge density in a metal or semiconductor, whose conduction electrons form an electron gas, creates long-range Coulomb fields which organize the system into collective motion[3]. SPPs have proved to be promising transducers in biological sensing[4, 5], and they are considered as a basis for a wide range of optical technologies, including modulators and switches[6] that can improve bandwidth over existing electronic devices, as well as for sub-wavelength amplifiers and lasers[7]. Interestingly, novel fundamental aspects of SPPs keep emerging[8-12], ranging from extraordinary momentum and spin in SPPs[8], to forces and torques produced by SPPs[13]. However, when it comes to enhancing optical chirality by SPPs the interest fades as the strict TM nature of surface plasmon plane waves dictates zero optical chirality[9, 10].

Here we show that SPPs launched by near field sources manifest rich and non-trivial chiral characteristics. We analytically solve for the optical chirality of such SPPs and explicitly show that this interesting behavior derives from the nature of these SPPs as intrinsic superposition states of surface plasmon plane waves. In particular, we demonstrate that the interference of plane wave modes in the near-field creates the observed chiral behavior of SPPs. We also introduce the concept of Natural Chiral Sources (NCS) that can launch chiral SPPs and result in enhanced net optical chirality. We show that naturally occurring chiral entities, such as helical DNA molecules, form templates for chiral sources that can launch surface plasmons carrying the chiral character of the source. This concept is of great significance for extending the spectral range of plasmonically enhanced chiroptical effects into the UV spectral range. This regime is of significant interest for chiral bio-sensing but has been inaccessible with the more common enhancement schemes based on localized surface plasmons.[14-16] Also, such NCSs may lay the groundwork for a new class of self-assembled chiral near-field sources that can mitigate the challenges associated with top-down nano-fabrication.

**Results**

Because of their coupled nature with photons, surface plasmons are light-like, $\omega_s(k) \approx c|k|$ for small $|k| \ll \omega/c$ and asymptotically approach a constant value $\omega_k(k) \approx \omega_p/\sqrt{1+\varepsilon_m}$ for $|k| \gg \omega/c$, where $\varepsilon_m$ is the permittivity of the metal embedded in vacuum. Unlike bulk plasmons, surface plasmons have components parallel and perpendicular to their two-dimensional propagation vector in the plane of the interface[17]. As long as the mean free path of electrons is much shorter than the wavelength of the plasmon, macroscopic electrodynamics can be utilized to describe the optically excited surface plasmons[1]. The interactions between near-field dipole sources and surface plasmons have been investigated in detail[17-26]. Here we will look into the chiral behavior of SPPs excited by differently polarized near-field sources in vicinity of a semi-infinite metal surface, See Figure 1a for instance. When a near-field source is placed close to an interface that can sustain SPPs, it can decay through different channels depending on its distance from the interface. At distances comparable to the emission wavelength, $d \approx \lambda/2$, the decay is primarily in the form of a photon[20]. In such a case, the fluorescent life-time shows an oscillatory behavior with distance d, which is due to the interference between the radiated and the reflected photons[20, 27]. At distances around d=50 nm, an optimum coupling of source photons and propagating surface plasmons is obtained[20, 28]. The coupling efficiency drops with decreasing $d$ and at very small distances, $d \approx 5$ nm, non-radiative processes dominate the decay-rate. These processes are determined by intrinsic loss-mechanisms in the metal, which include interband transitions, electron scattering loss and electron-hole excitation[20]. Apart from the distance of the source from the metal surface, its orientation plays an important role in determining its decay mechanism. A

vertical dipole couples only to the p-polarized reflected waves. However, an in-plane dipole can couple to both p-polarized and s-polarized components of the back-reflected waves. In the case of near-field excited surface charge density oscillations the fields due to the source dominate in the near-field, while the fields associated with propagating SPPs dominate for larger distances. The propagation of a p-polarized surface plasmon plane wave on a flat surface normal to z direction is described by $\boldsymbol{E} = E(z)\exp(i\boldsymbol{k}.\boldsymbol{\rho} - \omega t)$ where $\omega$ is the angular frequency, $\boldsymbol{k}$ is the wave vector which is parallel to the interface and $\rho$ is the radial variable in cylindrical coordinates. Surface modes excited by a near-field source possess, however, transverse as well as longitudinal components. In the case of a source placed close to an SPP-sustaining interface with the current density of $\boldsymbol{j}(\boldsymbol{r},t)$ the electric field of the SPP can be obtained by[24, 29]:

$$\boldsymbol{E}(\boldsymbol{r},t) = -\mu_0 \int dt' \int d^3\boldsymbol{r}' \overrightarrow{\boldsymbol{G}'}(\boldsymbol{r},\boldsymbol{r}',t-t') \partial \boldsymbol{j}(\boldsymbol{r},\boldsymbol{r}')/\partial t' \qquad (1)$$

where $\mu_0$ is the vacuum permeability and $\boldsymbol{G}'$ is the Fourier transform of the Green's tensor, $\vec{\boldsymbol{g}}(\boldsymbol{k},z,z',\omega)$, associated with the metal-dielectric infinite interface:

$$\boldsymbol{G}'(\boldsymbol{r},\boldsymbol{r}',t-t') = \int d^2\boldsymbol{k}/4\pi^2 \int d\omega/2\pi \, \vec{\boldsymbol{g}}(\boldsymbol{k},z,z',\omega) e^{i[k(r-r')-\omega(t-t')]} \qquad (2)$$

It is observed that $\boldsymbol{G}'$ has poles that are determined by the denominators of the Fresnel factors for p-polarized light. These poles relate to the surface plasmon modes and explain the resonant nature of these modes. If the metal slab has finite thickness, plasmon modes can be determined by the zeros of the transcendental equation: $1 + r_{1,2}^{(p)}(k_{sp}) r_{2,3}^{(p)}(k_{sp}) \exp(2ik_{2z}d) = 0$ where $r_{1,2}^{(p)}(k_{sp}), r_{2,3}^{(p)}(k_{sp})$ are the Fresnel reflection coefficients for p-polarized light for upper and lower media respectively[21]. The SPP wave-vector is determined by, $k_{sp} = k_0 \sqrt{(\varepsilon_1 \varepsilon_2)/(\varepsilon_1 + \varepsilon_2)}$ where $k_0$ is the free-space propagation constant and $\varepsilon_1$ and $\varepsilon_2$ are the permittivities of the surrounding medium and metal respectively. The vertical component of the wave-vector in medium $j$ is determined by $k_{jz} = \sqrt{k_j^2 - k_{sp}^2}$ with $\text{Im}\{k_{jz}\} > 0$. The condition $\text{Im}\{k_{jz}\} > 0$, selects the solutions on one Rieman sheet making the square root single valued[21]. Equation (1) can be used to find the explicit form of the electric field for a generic orientation of dipole placed at a distance d above a flat SPP-sustaining surface. To keep it intuitive, we first evaluate the special case of a z-polarized dipole at a distance d above the metal slab. Surface plasmons can be excited by a vertical as well as a horizontal electric dipole as both orientations have a p-polarized component that couples to the longitudinal electron density oscillations. This is not the case for vertical magnetic dipoles due to their lack of any p-polarized components[30].

**Optical Chirality from SPPs excited by a linearly polarized dipole.** Solving for the Green's tensor of a dielectric-metal interface with a time-harmonic source the electric field of a surface plasmon due to a z-polarized dipole can be written, in cylindrical coordinate, as[21, 23, 24, 29]:

$$\boldsymbol{E} = 2\Re\{\frac{M(k_{sp},\omega_0)}{\varepsilon_0}(-i)\frac{k_{sp}}{k_{z1}}p_0 e^{ik_{z1}d}e^{ik_{z1}z}e^{-i\omega t}[H_1^{(1)}(k_{sp}\rho)\hat{\rho} + i\frac{k_{sp}}{k_{z1}}H_0^{(1)}(k_{sp}\rho)\hat{z}]\} \qquad (3)$$

where $M(k_{sp},\omega_0) = \dfrac{-k_{z1}k_{z2}}{4}\dfrac{k_{z1}\varepsilon_2 - k_{z2}\varepsilon_1}{\varepsilon_1^2 - \varepsilon_2^2}$, $k_{sp}$ is the SPP wave-vector, $k_{z1}, k_{z2}$ are the vertical components of the wave-vector in air and metal respectively, $\varepsilon_0$ is the vacuum permittivity and $H_0^1(k_{sp}\rho)$ and $H_1^1(k_{sp}\rho)$ are different orders of complex Hankel Functions of first kind. It is physically intuitive that such an electric field should be azimuthally symmetric as the dipole source dictates such symmetry. A closer look at equation (3) reveals that such an SPP mode preserves its TM nature due to independence of the electric field from the azimuthal angle, which in turn deprives the magnetic field of any longitudinal component and leaves the magnetic field only with an azimuthal component. As a result, given the definition of optical chirality density $C$, as[31],

$$C \equiv \frac{\varepsilon_0}{2} \boldsymbol{E} \cdot \nabla \times \boldsymbol{E} + \frac{1}{2\mu_0} \boldsymbol{B} \cdot \nabla \times \boldsymbol{B} \tag{4}$$

where $\boldsymbol{E}$ and $\boldsymbol{B}$ are the time-dependent electric and magnetic fields, one finds that $C$ unequivocally vanishes for such surface plasmon modes. The result comes as no surprise as the optical chirality density is a pseudo-scalar, being anti-symmetric under parity transformation, while the physical system at hand is invariant under parity. One may think that the presence of the substrate has already broken the z→-z symmetry. However, one should consider the combination of the substrate, the dipole and the image dipole, which together respect the parity symmetry (Also see Supporting Note 1). The situation is, however, starkly different for a point source that breaks the azimuthal symmetry. For instance an in-plane x-polarized linear dipole launches surface plasmons with cosφ symmetry. This is because it is the longitudinal components of the wave vector that couple to the surface plasmons and they dominate mostly in the direction of the dipole. The electric field of a surface plasmon due to an x-polarized dipole source can be written as[23, 24, 29]

$$\boldsymbol{E} = \frac{M(k_{sp},\omega_0)}{\varepsilon_0} p_0 e^{ik_{z1}d} e^{ik_{z1}z} e^{-i\omega t} \{[H_0^{(1)}(k_{sp}\rho) - \frac{1}{k_{sp}\rho} H_1^{(1)}(k_{sp}\rho)]\cos\varphi \hat{\rho} - \frac{1}{k_{sp}\rho} H_1^{(1)}(k_{sp}\rho)\sin\varphi \hat{\varphi} - \frac{ik_{sp}}{k_{z1}} H_1^{(1)}(k_{sp}\rho)\cos\varphi \hat{z}\} \tag{5}$$

The explicit form of the electric field shows that such an SPP ceases to be transverse through incorporating an anti-symmetric transverse component in its electric field that in turn results in a longitudinal component of the magnetic field. This is where the first condition of obtaining non-trivial values for optical chirality is met, as we make sure that such SPP fields will have parallel components of electric and magnetic fields. The second requirement is the existence of a phase shift between such parallel components. This requirement is inherent to electric and magnetic fields, at least in the near-field. Making use of Maxwell-Faraday equation, we derive the magnetic field of the SPP (see Supporting Note 2) and then we use equation (4) to calculate the explicit form of the optical chirality to be

$$C = \frac{|\zeta|^2 \sin 2\varphi}{2\rho\omega} \Im\{H_1^{(1)}(H_0^{(1)*} - \frac{1}{k_{sp}\rho} H_1^{(1)*})(\frac{k_{sp}}{k_{z1}} + \frac{k_{z1}}{k_{sp}}) - \frac{H_1^{(1)*}}{k_{sp}^*}(H_0^{(1)} - \frac{1}{k_{sp}\rho} H_1^{(1)})(k_{z1} + \frac{k_{sp}^2}{k_{z1}})\} \quad (6)$$

where $\zeta = \frac{M(k_{sp}, \omega_0)}{2\varepsilon_0} p_0 e^{ik_{z1}d}$. A more expanded and explicit form of the optical chirality can be obtained by substituting $k_{sp} = \Re(k_{sp}) + i\Im(k_{sp}), k_{z1} = \Re(k_{z1}) + i\Im(k_{z1})$, where $\Re(k_{sp}), \Im(k_{sp})$ are the real and imaginary parts of the in-plane SPP wave-vector and $\Re(k_{z1}), \Im(k_{z1})$ are the real and imaginary parts of the vertical SPP wave-vector. Also we could substitute $H_{0,1}^{(1)}(k_{sp}\rho) = J_{0,1}^{(1)}(k_{sp}\rho) + iY_{0,1}^{(1)}(k_{sp}\rho)$, where $J_{0,1}^{(1)}(k_{sp}\rho), Y_{0,1}^{(1)}(k_{sp}\rho)$ are cylindrical Bessel functions of the first and second kind, respectively. The permittivity of silver at λ=357 nm is $\varepsilon_r$=-2.145 and $\varepsilon_i$ =0.275[32], where $\varepsilon_r$ and $\varepsilon_i$ are the real and imaginary parts of the permittivity, respectively. Given that $k_{sp} = k_0\sqrt{(\varepsilon_1\varepsilon_2)/(\varepsilon_1+\varepsilon_2)}$ and $k_{1z} = \sqrt{k_1^2 - k_{sp}^2}$, one obtains $\Re k_{sp} = 1.353k_0, \Im k_{sp} = .07k_0$ and $k_{z1} = i0.914k_0$, also $\Im k_{sp} \ll |k_{sp}|$. Substituting these values one can simplify equation (6) further:

$$C = \frac{|\zeta|^2 \sin 2\varphi}{2\rho\omega}\{0.305[Y_1^{(1)}J_0^{(1)} - Y_0^{(1)}J_1^{(1)}] - 1.599[J_1^{(1)}J_0^{(1)} + Y_1^{(1)}Y_0^{(1)} - \frac{0.712}{\rho k_0}(|J_1^{(1)}|^2 + |Y_1^{(1)}|^2)]\} \quad (7)$$

where Y and J are functions of $k_{sp}\rho$ (For details of the derivation see Supporting Note 2). A closer look at equations (6) and (7) brings several interesting points to attention. Strikingly, contrary to the common perception of surface plasmons exhibiting zero density of optical chirality we observe an absolutely non-trivial behavior of surface plasmons in terms of optical chirality. First of all, the optical chirality for a SPP excited by an in-plane linearly polarized dipole (LSPP) is locally strictly non-zero even at locations remote from the excitation source. This contrasts the fields of a linearly polarized dipole that deliver zero optical chirality, except in the immediate near-field of the dipole where the electric and magnetic fields are yet to be completely transverse. In the case of a point-like dipole source such spatial extension approaches ultra-subwavelength values. The reason for the non-trivial behavior of optical chirality of such SPPs lies in the fact that such surface plasmon modes are superpositions of plasmon plane wave modes that by themselves render zero optical chirality but in a superposition state can interfere to produce a locally non-zero optical chirality. Recently, such near-field interference effects have been utilized for the directional launching of SPPs[33]. The dependence of surface plasmon's optical chirality density on the azimuthal angle, as shown in equation (6), is anti-symmetric as expected for a quantity that is anti-symmetric under parity. The decay of the optical chirality in the direction of propagation of the surface plasmon depends in a complex fashion on cylindrical Bessel functions, which rapidly decay as the wave propagates outward from the center. This interesting behavior loses some of its significance, however, if we are interested merely in the net behavior of the optical chirality. The anti-symmetric azimuthal dependence of the optical chirality density ensures that integration over the entire surface yields no net chirality. This behavior is certainly dissatisfactory from the biosensing point of view as it does not alleviate the discrimination between the enantiomers of a racemic mixture.

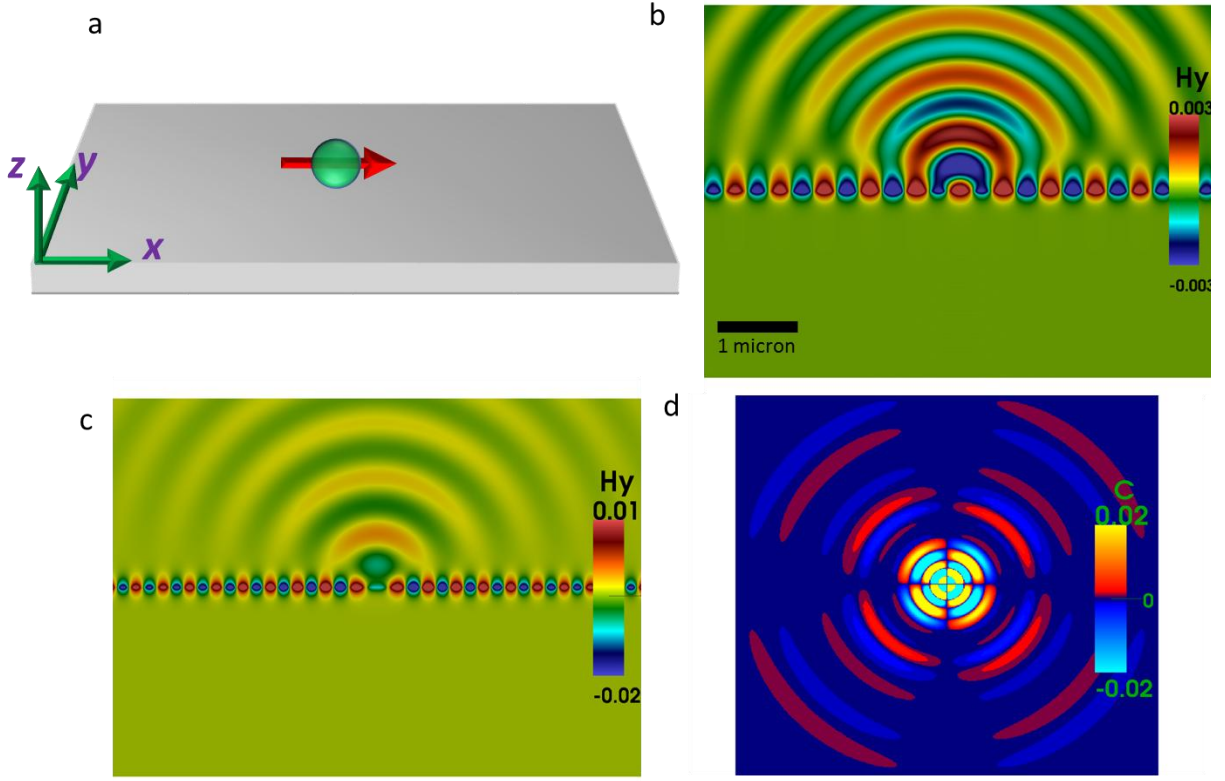

**Figure 1.** (a) Schematic of the system of interest. Photons from a near-field dipole couple to surface charge density of the silver slab. (b) Simulated transverse magnetic field component, H$_y$, of the excited SPP at 382 nm. Although SPPs are excited, the coupling is not ideal and a considerable portion of the EM energy of the dipole propagates to the far-field. (c) The excitation of SPPs for an infinitely thick silver slab is most efficient at λ=357 nm. (d) Numerically calculated optical chirality of the LSPP. It exhibits the analytically predicted sin2φ dependence. The scale bar of Figure 1b applies to all the graphs.

To gain further insight into the behavior of optical chirality in the aforementioned circumstances we did a series of vectorial full-wave simulations. A dipole source with electric dipole moment of 1 D (1 D=3.3356×$10^{-30}$ C.m) was placed 50 nm above a thick silver slab in x direction. The infinite thickness of the slab ensures the absence of any reflected waves from the lower interface and also inhibits Fabry-Perot type modes in the metal. Due to the lack of back-reflected waves, the observed fields in the simulated plane are solely due to the exciting source. The specific vertical height of 50 nm was chosen to represent a real-world position of a near-field source. Also in closer distances dissipative channels hinder efficient coupling of dipole near-field photons with the surface modes (For the details of the simulations See Methods). Figure 1b and 1c show the transverse component of the magnetic field, $H_y$, in xz plane. It is noticed that as the frequency of the source approaches that of the SPP resonance, the surface modes couple more efficiently to the source. In the investigated geometry the SPP resonance occurs at around 357 nm, Figure 1c. We evaluated the optical chirality density in a plane 5nm above the silver surface and plotted $C$ for λ=357 nm (Figure 1d). The plot confirms that the simulated optical chirality behaves as predicted by equation (6). $C$ vanishes in the direction of the dipole oscillations as well as perpendicular to it. It also rapidly decays as propagates away from the center. It should be mentioned, however, that despite this rapid decline, such SPP shows a spatial extent of locally enhanced chiral fields which is comparable to schemes based on localized surface plasmons[34].

**Optical Chirality from SPPs excited by a circularly polarized dipole.** The observed chiral behavior is totally distinct when we consider the surface plasmons excited by a circularly polarized dipole: $\mathbf{p} = p_0(\hat{i} + e^{i\pi/2}\hat{j})$, where $\hat{i}, \hat{j}$ are unit vectors in x and y directions, respectively. Such a near-field source has recently attracted attention both experimentally[33, 35] and theoretically[36]. Although the metal slab is invariant under xy-plane mirror reflection, exciting the surface plasmons with a circularly polarized dipole which itself is chiral breaks the reflection symmetry and bestows the physical system with chirality. As a consequence, one expects a chiral behavior that differs from the one described above for the linearly polarized dipole excitation. A schematic drawing of the physical system is shown in Figure 2a. The transverse magnetic field in the xy plane, $H_y$, is plotted in Figure 2b. It is evident that the excited surface plasmon exhibits a circular character due to the exciting circularly polarized dipole. This magnetic field pattern results from the superposition of the surface plasmons launched normal to each other with a $\pi/2$ phase shift. The same circular nature holds true for the electric field of the surface plasmon. The electric field of a circular surface plasmon polariton, CSPP, is calculated by superposing the electric fields of x-polarized and y-polarized dipoles with a $\pi/2$ phase shift. The explicit form of the electric field can be calculated to be:

$$\boldsymbol{E} = \frac{M(k_{sp},\omega_0)}{\varepsilon_0} p_0 e^{ik_{z1}d} e^{ik_{z1}z} e^{-i\omega t} e^{i\varphi} \{[H_0^{(1)}(k_{sp}\rho) - \frac{1}{k_{sp}\rho}H_1^{(1)}(k_{sp}\rho)]\hat{\rho} - \frac{i}{k_{sp}\rho}H_1^{(1)}(k_{sp}\rho)\hat{\varphi} - \frac{ik_{sp}}{k_{z1}}H_1^{(1)}(k_{sp}\rho)\hat{z}\} \quad (8)$$

The electric field of the SPP carries an $e^{i\varphi}$ dependence with both symmetric and anti-symmetric parts. The polarization state of such a SPP is interesting, as one immediately notices that the SPP has both longitudinal and transverse components. This behavior is absent in a strictly TM polarized surface plasmon plane wave. The transverse component of the electric field has an intrinsic phase shift relative to the longitudinal component and decays on a length-scale characterized by $k_{sp}\rho$. We make use of the Maxwell-Faraday equation to calculate the magnetic field (See Supporting Note 3) and by the same token calculate the optical chirality density as:

$$C = \frac{|\zeta|^2}{2\rho\omega} \Re\{-|H_1^{(1)}|^2 [\frac{1}{|k_{sp}|^2 \rho^2} + \frac{k_{z1}}{|k_{sp}|^2 \rho} + \frac{2k_{sp}^*}{k_{z1}^* k_{sp}\rho}]$$
$$+ H_1^{(1)} H_0^{(1)*} [\frac{-k_{sp}}{k_{z1}} + \frac{1}{k_{sp}\rho}] \quad (9)$$
$$+ H_1^{(1)*} H_0^{(1)} [\frac{k_{z1}}{k_{sp}^*} + \frac{k_{sp}^2}{k_{sp}^* k_{z1}} + \frac{2k_{sp}^*}{k_{z1}^*}]\}$$

As expected any φ dependence vanishes in the optical chirality density, which promises an azimuthally symmetric behavior. Also, it decays as the SPP travels away from the center. However, it keeps its oscillatory character due to the cylindrical Bessel functions. This oscillatory behavior is a result of an interplay between terms with different orders of the Bessel functions. For the more explicit form of the optical chirality density see Supporting Note 3. We tested the predicted behavior of the optical chirality density of a chiral SPP by vectorial full-wave simulations. A circularly polarized dipole was used to launch the CSPP. Such a dipole was produced by superposing two normal dipoles, each with a dipole moment of 1 D (1 D=$3.3356 \times 10^{-30}$ C.m), which oscillated with a $\pi/2$ phase shift. The electromagnetic fields were simulated in a plane 5 nm above the surface of the silver slab and subsequently the optical chirality was calculated from the collected fields and was plotted as seen in Fig 2c. Since the optical chirality of the source itself is non-zero, the optical chirality density of the CSPP was normalized to that of the source. The points with zero optical chirality of the source were discarded to avoid diverging results. The simulations match very well with the analytical calculations. We note that the optical chirality density decays slower than that of a LSPP. Also locally one expects to get a non-zero net optical chirality within a full period of the azimuthal angle and a radial range of square root of wavelength of the surface plasmon, which is roughly the period of the optical chirality alternation. This relation is a direct consequence of the radial dependence of the electromagnetic fields on the first power of the cylindrical Bessel functions and of the optical chirality on the second power of these functions. If a biological sample of interest is confined to such an area of uniform sign of optical chirality, which is microns in perimeter and hundreds of nanometers in radial extension (see Figure 2c), an excess of chiral interaction between one enantiomer and the chiral SPP is obtained. Due to the non-zero net chiral

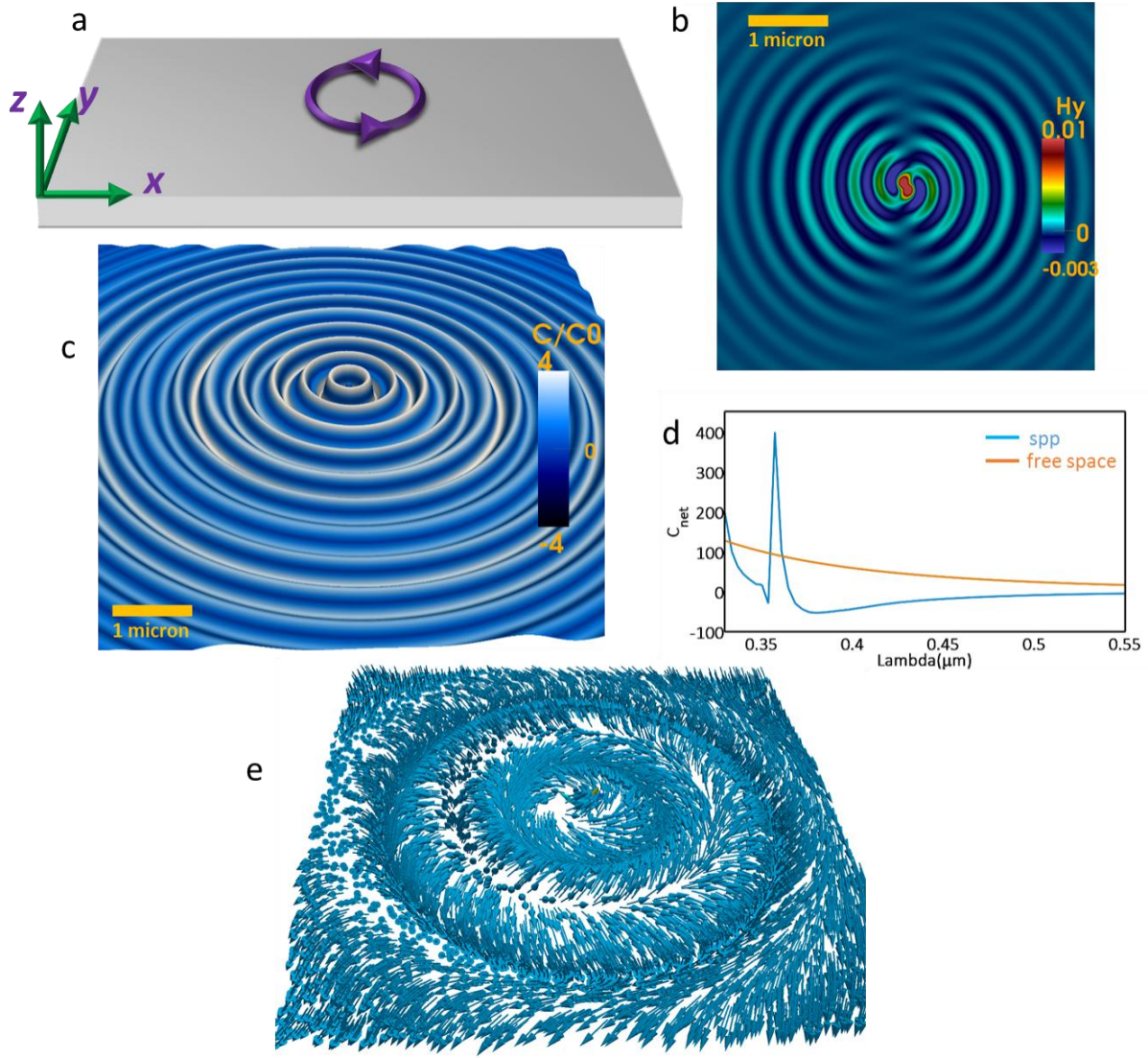

**Figure 2.** (a) Schematic overview of the simulation set-up. The dipole is placed 50 nm above the silver slab (b) Simulated $H_y$ of the excited SPP at resonance, $\lambda=357$ nm. The SPP carries the circular characteristic of the exciting field. (c) Numerically simulated optical chirality of the SPP at $\lambda=357$ nm. (d) Calculated total optical chirality in a surface 5nm above the interface. (e) The behavior of the optical chirality flow is drastically different than that of the SPP momentum. It is closely related to the transverse spin angular momentum of the SPP. The scale bar of Figure 2b applies to 2e as well.

enhancement, even in the absence of any confinement, we expect an excess in the chiral interaction between the light and one of the racemic enantiomers. Even in the absence of any confinement we would expect a net enhancement of optical chirality as the SPP inherits the chiral character of the source. To illustrate this point we integrated the optical chirality density over the entire simulation surface and plotted the results versus the wavelength (Figure 2d). One can clearly see that away from the plasmon resonance the only contribution comes from the dipole source, which itself is chiral and should exhibit total non-zero chirality. However, as we approach the wavelengths at which the optimal coupling of the light and surface plasmons occur, the net optical chirality increases dramatically, which is due to the contribution from the SPP. This result confirms the ability to enhance optical chirality through locally excited SPPs. Importantly, in the case of CSPPs one obtains not only a local enhancement of chiral fields, but also achieves a net enhancement of the optical chirality over the entire surface, paving the way for chiral sensing schemes based on SPPs. Moreover, it is worth mentioning that the wavelengths where such enhanced effects occur lie primarily in the UV region of the spectrum, which is inaccessible with localized plasmon resonances of common plasmonic structures based on gold and silver. This spectral range is, however, the relevant range for enhancing electronic Circular Dichroism (CD) spectroscopy.[37] In addition to the pseudo-scalar optical chirality density, a flow of chirality can be defined for an electromagnetic wave, $\boldsymbol{F} \equiv \{\boldsymbol{E} \times (\nabla \times \boldsymbol{B}) - \boldsymbol{B} \times (\nabla \times \boldsymbol{E})\}/2$, which together with the optical chirality satisfy the continuity equation[31, 38],

$\nabla \cdot \boldsymbol{F} + \partial C / \partial t = -1/2(\boldsymbol{j} \cdot \nabla \times \boldsymbol{E} + \boldsymbol{E} \cdot \nabla \times \boldsymbol{j})$, where $\boldsymbol{j}$ is the current density. Although the optical chirality for a plane wave SPP is strictly zero, the chirality flow is non-zero and it can be shown to be proportional to the transverse spin angular momentum of the SPP[8, 9]. This gains special significance as it implies that chirality flow can lead to opposite forces on molecules of different handedness[39]. Such handedness-dependent forces have been investigated and interpreted through different approaches[8, 11, 13, 39, 40]. The chirality flow of the CSPP is shown in Figure 2e. It is evident that the chirality flow is dominantly azimuthal, mostly perpendicular to the direction of the energy flow and linear momentum of the CSPP. Such a transverse flow would exert an unconventional and extraordinary force on a chiral molecule perpendicular to the conventional pushing and gradient forces, which are mainly radial.

**Chiral SPPs generated by natural chiral sources.** Despite recent advances in the design and realization of near field sources capable of launching circular SPPs, such sources are not readily available in nature and their experimental realization remains challenging[33, 35, 41]. In this section we show that a chiral molecule can mimic the chiral behavior of a circularly polarized near field source. Specifically, we argue that chiral molecules are capable of launching SPPs with a non-trivial chiral character. One example of a chiral molecule is given by a DNA-like helix that can be either left-handed or right-handed depending on whether the electric and magnetic dipoles are parallel or anti-parallel. Due to the geometrical chiral character of such combination of electric-magnetic dipoles, one expects to obtain non-zero optical chirality from EM fields of these dipoles in free space. The optical chirality for a combination of arbitrarily oriented electric and magnetic dipoles is given by (see Supporting Note 4):

$$C = \frac{\omega}{8\pi^2 c^2 \varepsilon_0} \text{Im}\{\frac{k^4}{r^2}[\vec{p}.\vec{m} - (\vec{n}.\vec{p})(\vec{n}.\vec{m})] - \frac{2k^2}{r^4}(\vec{p}.\vec{m}) + \frac{2k^2}{r^4}(\vec{n}.\vec{p})(\vec{n}.\vec{m}) + (\frac{1}{r^6} + \frac{k^2}{r^4})[\vec{p}.\vec{m} + 3(\vec{n}.\vec{p})(\vec{n}.\vec{m})]\} \quad (10)$$

It is readily deduced from this expression that the optical chirality directly depends on the relative orientation of the dipoles and no matter what the orientation is, in the absence of a phase difference between the electric and magnetic dipoles, the optical chirality is inevitably zero. This is due to the fact that without a phase term, the expression inside the brackets will lack any imaginary part. Of course, however, a phase shift of π does not meet this criterion. Artificial analogues of such natural chiral structures can be synthesized using bottom-up fabrication techniques that utilize chiral molecules as templates for chiral metal structures that support free electron motion. Metallization of DNAs is an example of such techniques[42-47]. When it comes to the radiation of these chiral dipoles near a metallic surface, such a combination of electric and magnetic dipoles will launch surface plasmons normal to one another. More interestingly, due to the intrinsic $\pi/2$ phase shift between the electric and magnetic dipoles in the helix, the SPPs excited by these dipoles carry the same phase information. The resulting SPP corresponds to a SPP launched by two perpendicular electric dipoles with $\pi/2$ phase shift. This intuitive picture was validated by numerical simulations, in which surface plasmons were excited by a right-handed helix (for details of the simulations see Methods). One expects the z component of the electric field to follow the pattern of a CSPP, as is seen in Figure 3b. The fields of the surface plasmons launched by these two dipoles dominate in perpendicular angles, see Supporting Figure.1. In the next step the optical chirality density was evaluated in a surface 5nm above the silver slab and then was normalized to that of the source (Figure 3c). The points with zero optical chirality of the source were discarded to avoid diverging results. The optical chirality distribution resulting from NCS excitation is dramatically different from those obtained with individual electric or magnetic dipoles. For a chiral dipole with non-dominant electric or magnetic components, the optical chirality density is azimuthally symmetric and decays radially. The visible asymmetry of the optical chirality along the y direction in Figure 3c, is due to a dominant component of SPP launched by the magnetic dipole. Interestingly, in the immediate near-field the optical chirality of the source fields is large enough to curb any dominance by that of the SPP. As we go further away from the source, however, the SPP fields and subsequently its optical chirality dominate. Figure 3c shows that the produced optical chirality is of similar magnitude as observed before for a CSPP, so naturally we expect to see the total optical chirality to behave similarly as that of the CSPP. The total optical chirality, obtained by integration over the entire surface, is plotted *vs*. the excitation wavelength in Figure 3d. Far from the resonance the electromagnetic fields are dominated by those of the source and the net optical chirality is non-zero but small. However, in the frequency window where optimal excitation of the SPP occurs, the net optical chirality rises dramatically. Right after the resonance peak this value becomes negative, as the phase between the electric and magnetic fields reverses when the wavelength is scanned across the resonance. We emphasize that the finding that optical chirality can be generated by NCS, which may by readily available in nature, is intriguing and non-trivial.

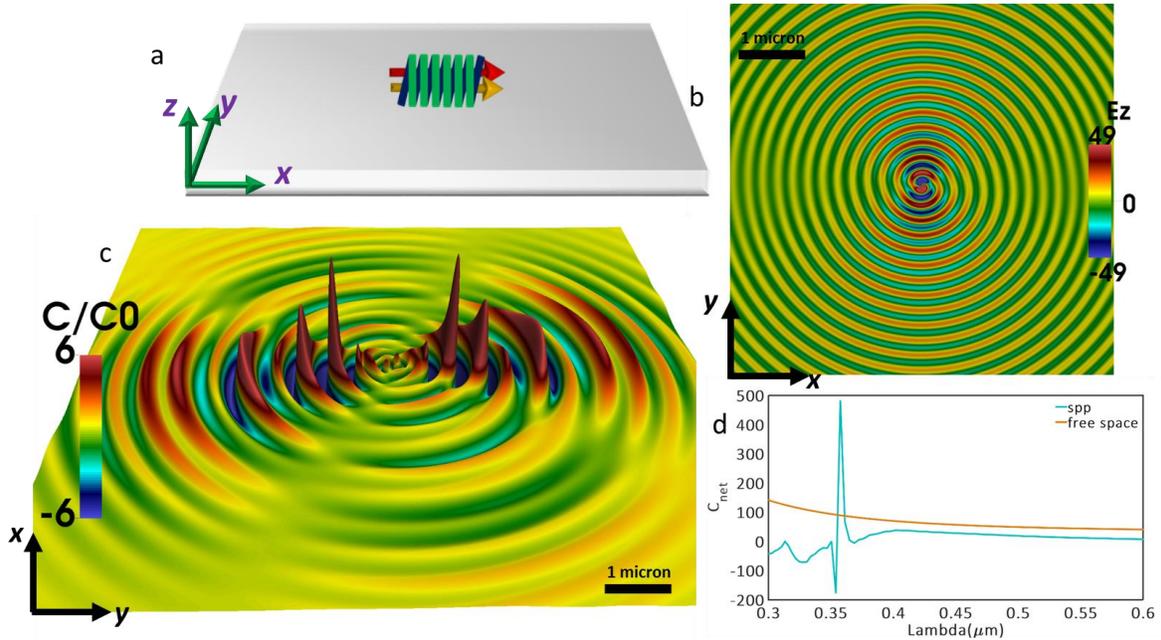

**Figure 3.** (a) The chiral molecule is represented by a pair of parallel electric and magnetic dipoles with strengths of $3.3356\times10^{-30}$ C.m and $197.096\times10^{-24}$ J.T$^{-1}$ (b) The simulated $E_z$ of the excited SPP at $\lambda=357$ nm. Despite the linear polarization of the source the SPP exhibits a circular character(c) The numerically calculated optical chirality of the SPP. It shows a pattern similar to that of a CSPP. The scale bar has the same dimensions as in 3b. (d) Calculated total optical chirality of on a surface 5nm above the interface.

**Discussion**

In summary, we investigated enhancement of optical chirality through surface plasmon polaritons. We showed that locally excited SPPs can exhibit enhanced optical chirality both locally and globally, depending on the polarization of the exciting near-field source. Such locally excited surface plasmons are inherently superposition of p-polarized surface plasmon plane waves, and due to the near-field interference of such modes SPPs with non-trivial chiral characters can emerge. We demonstrated both analytically and numerically that for an in-plane exciting dipole source SPPs with locally enhanced optical chirality can emerge. However, due to the anti-symmetric nature of the optical chirality of such SPPs the net enhancement of optical chirality is zero. Importantly, we further showed that for SPPs excited by a chiral source not only locally enhanced optical chirality but also a non-vanishing net enhancement over extended areas can be obtained. This enhancement can reach manifold values upon approaching the resonance frequency of the SPP. We also introduced the concept of Natural Chiral Sources, which circumvent the challenges associated with realization of a chiral near-field source through experimental procedures. In particular, we demonstrated that SPPs generated by a chiral molecule such as a DNA-like helical molecule or a metallized DNA carry along chiral characters similar to those of a CSPP. We attributed this effect to the superposition of SPPs excited by the chiral molecule, which mimic a pair of parallel or anti-parallel electric and magnetic dipoles with intrinsic phase differences. These findings can be of great interest in applications related to chiral bio-sensing. From an experimental point of view one should keep in mind that the optical chirality density of SPPs will experience a rate of exponential decay in vertical direction double that of the electric and magnetic fields. This is of significance in that such a density will be highly confined to the metal surface and to its vertical immediate vicinity. This confinement makes it possible to selectively measure chiral bio-molecules on or near the metal substrate. In combination with selective capture chemistries, this can pave the way to selective chiral detection schemes. Furthermore, a rapid decay of the optical chirality along the vertical direction creates a large gradient of optical chirality density and, consequently, may lead to distinct chiral forces.[14] Finally, the ability to move the spectral range of operation of plasmonically enhanced chiral sensing into the UV part of the electromagnetic spectrum can contribute to improving the structural characterization of biomolecules. We should emphasize that the chiral bio-sensing schemes based on SPPs would be surface-dependent and differ from conventional techniques based on transmitted light.

**Methods**

**Numerical Simulations**

Full-wave Finite –Difference Time-Domain simulations were used for the simulations. In all the simulations silver was used as the metal medium. The optical constant were extracted from Johnson and Christy[32]. The dipole was positioned at 50 nm above the metal surface and the data were collected from a surface 5nm above the surface. In the simulations related to NCSs the electric dipole was chosen to be 1 D, (1 D=$3.3356\times10^{-30}$ C.m) and the magnetic dipole moment was 4 $\mu_B$ where $\mu_B$ is the Bohr magneton ($\mu_B$= $49.274\times10^{-24}$ J.T$^{-1}$). The reason for introducing the slight dominance of the magnetic character was to mimic the electric and magnetic dipoles of a helix. The calculated magnetic field is for a single current loop. A helix, however, consists of several turns or loops which justifies multiplying the magnetic moment with a number which we chose to be 4. In all the simulations fine meshes with maximum mesh size of $\lambda/70$ was used.

**Author Information**

Corresponding Authors:

E-mail: halizade@bu.edu (M.H.A.).

E-mail: bmr@bu.edu (B.M.R.).

**Acknowledgement**

This work was supported by the U.S. Department of Energy, Office of Basic Energy Sciences, Division of Materials Science and Engineering under Award DOE DE-SC0010679.

# Supporting Information: Enhanced Optical Chirality through Locally Excited Surface Plasmon Polaritons

M.H. Alizadeh[‡,¶] and Björn M. Reinhard[†,¶]

[‡]Department of Physics, [†]Department of Chemistry, and [¶]The Photonics Center,

Boston University, Boston, MA 02215, United States

**Supporting Note 1.**

The optical chirality for a generic orientation of a dipole can be calculated to be:

$$C = \frac{M(k_{sp},\omega_0)}{\varepsilon_0} p_0 e^{ik_{z1}d} e^{ik_{z1}z} e^{-i\omega t} \{[\frac{-ik^*_{spp}}{k^*_{z1}} P_z^* H_1^{(1)*}(k_{spp}\rho) + (H_0^{(1)*} - \frac{1}{k^*_{spp}\rho} H_1^{(1)*})(P_x^* \cos\phi + P_y^* \sin\phi)]$$

$$[\frac{-i}{\rho\omega}((-P_x \sin\phi + P_y \cos\phi)(\frac{-ik_{spp}}{k_{z1}} P_z^* H_1^{(1)}))]\}$$

$$+\{[\frac{iH_1^{(1)*}}{\omega k^*_{spp}\rho}(P_x^* \sin\phi - P_y^* \cos\phi)][P_z(\frac{k_{spp}}{k_{z1}})^2 \frac{\partial H_0^{(1)}}{\partial \rho} + \frac{ik_{spp}}{k_{z1}} \frac{\partial H_1^{(1)}}{\partial \rho}(P_x \cos\phi + P_y \sin\phi)]\} \quad (1\text{-}1)$$

$$+\{(\frac{-i}{\omega})[-(\frac{k_{spp}}{k_{z1}})^2 P_z^* H_0^{(1)*} - \frac{ik^*_{spp}}{k^*_{z1}} H_1^{(1)*}(P_x^* \cos\phi + P_y^* \sin\phi)]$$

$$[\frac{1}{\rho}[(\frac{1}{k_{spp}} \frac{\partial H_1^{(1)}}{\partial \rho}(-P_x \sin\phi + P_y \cos\phi) - (H_0^{(1)} - \frac{1}{k_{spp}\rho} H_1^{(1)})(P_y \cos\phi - P_x \sin\phi)]]\}$$

Despite the complexity of equation (1-1), it is not formidable to note that when a dipole possesses only a z-component, all the terms in the optical chirality go to zero.

**Supporting Note 2.**

The magnetic field of the SPPs, was calculated suing Maxwell-Faraday Equation in cylindrical coordinates.

$$B_\rho = \frac{1}{i\omega}[\frac{\partial E_z}{\rho\partial\varphi} - \frac{\partial E_\varphi}{\partial z}],$$

$$B_\varphi = \frac{1}{i\omega}[\frac{\partial E_\rho}{\partial z} - \frac{\partial E_z}{\partial \rho}], \quad (2\text{-}1)$$

$$B_z = \frac{1}{i\rho\omega}[\frac{\partial(\rho E_\varphi)}{\partial \rho} - \frac{\partial E_\rho}{\partial \varphi}]$$

Then the optical chirality was obtained using equation (4). Since all the calculations were done for time-harmonic fields, the Fourier transform of equation (4), $C = \text{Im}[\boldsymbol{E}^*.\boldsymbol{B}]$, was used instead. The magnetic field was calculated for different polarizations of the exciting dipole. For the case of x-polarized dipole, the magnetic field is calculated to be:

$$\boldsymbol{B} = \frac{M(k_{sp},\omega_0)}{\varepsilon_0} p_0 e^{ik_{z1}d} e^{ik_{z1}z} e^{-i\omega t} \{[\frac{k_{sp}}{k_{z1}} + \frac{k_{z1}}{k_{sp}}]\frac{H_1^{(1)}(k_{sp}\rho)}{\rho\omega}\sin\varphi\hat{\rho} + \frac{1}{\omega}[(H_0^{(1)}(k_{sp}\rho)$$

$$-\frac{1}{k_{sp}\rho}H_1^{(1)}(k_{sp}\rho))k_{z1} + \frac{k_{sp}^2}{k_{z1}}\frac{\partial H_1^{(1)}(k_{sp}\rho)}{\partial(k_{sp}\rho)}]\sin\varphi\hat{\varphi} \quad (2\text{-}2)$$

$$-[(H_0^{(1)}(k_{sp}\rho) - \frac{1}{k_{sp}\rho}H_1^{(1)}(k_{sp}\rho)) - \frac{\partial H_1^{(1)}(k_{sp}\rho)}{\partial(k_{sp}\rho)}]\frac{\sin\varphi}{\rho\omega}\hat{z}\}$$

from which the optical chirality can be calculated to be

$$C = \frac{|\zeta|^2 \sin 2\varphi}{2\rho\omega} \Im\{H_1^{(1)}(H_0^{(1)*} - \frac{1}{k_{sp}\rho}H_1^{(1)*})(\frac{k_{sp}}{k_{z1}} + \frac{k_{z1}}{k_{sp}}) - \frac{H_1^{(1)*}}{k_{sp}^*}(H_0^{(1)} - \frac{1}{k_{sp}\rho}H_1^{(1)})(k_{z1} + \frac{k_{sp}^2}{k_{z1}})\} \quad (2\text{-}3)$$

where $\zeta = \frac{M(k_{sp},\omega_0)}{2\varepsilon_0} p_0 e^{ik_{z1}d}$. Plugging in for real and imaginary parts of the surface plasmon wave vector $k_{sp} = \Re(k_{sp}) + i\Im(k_{sp}), k_{z1} = \Re(k_{z1}) + i\Im(k_{z1})$ we get:

$$C = \frac{|\zeta|^2 \sin 2\varphi}{2\rho\omega} \{ \frac{2\Re k_{sp}}{|k_{sp}|^2} [\Re k_{z1}(1 + \frac{\Re k_{sp}^2 - \Im k_{sp}^2}{|k_{z1}|^2}) + \frac{2\Re k_{sp} \Im k_{sp} \Im k_{z1}}{|k_{z1}|^2}][Y_1^{(1)} J_0^{(1)} - Y_0^{(1)} J_1^{(1)} + \frac{\Im k_{sp}}{|k_{sp}|^2 \rho}(|J_1^{(1)}|^2 + |Y_1^{(1)}|^2)] +$$

$$+ \frac{2\Re k_{sp}}{|k_{sp}|^2}[\Im k_{z1}(1 - \frac{\Re k_{sp}^2 - \Im k_{sp}^2}{|k_{z1}|^2}) + \frac{2\Re k_{sp} \Im k_{sp} \Re k_{z1}}{|k_{z1}|^2}][J_1^{(1)} J_0^{(1)} + Y_1^{(1)} Y_0^{(1)} - \frac{\Re k_{sp}}{|k_{sp}|^2 \rho}(|J_1^{(1)}|^2 + |Y_1^{(1)}|^2)]\}$$

(2-4)

Y and J Bessel functions are functions of $(k_{sp}\rho)$ which has been dropped for brevity. Given that $k_{sp} = k_0\sqrt{(\varepsilon_1\varepsilon_2)/(\varepsilon_1 + \varepsilon_2)}$ and $k_{1z} = \sqrt{k_1^2 - k_{sp}^2}$, where $k_0$ is the propagation constant in vacuum, on resonance $k_{1z}$ becomes purely imaginary, leaving the real part of the vertical wave-vector zero. This simplifies equation (2-4) to:

$$C = \frac{|\zeta|^2 \sin 2\varphi}{2\rho\omega} \{ \frac{2\Re(k_{sp})}{|k_{sp}|^2}[\frac{2\Re k_{sp} \Im k_{sp} \Im k_{z1}}{|k_{z1}|^2}]$$

$$[Y_1^{(1)} J_0^{(1)} - Y_0^{(1)} J_1^{(1)} + \frac{\Im k_{sp}}{|k_{sp}|^2 \rho}(|J_1^{(1)}|^2 + |Y_1^{(1)}|^2)] + \qquad (2\text{-}5)$$

$$\frac{2\Re(k_{sp})}{|k_{sp}|^2}[\Im k_{z1}(1 - \frac{\Re k_{sp}^2 - \Im k_{sp}^2}{|k_{z1}|^2})][J_1^{(1)} J_0^{(1)} + Y_1^{(1)} Y_0^{(1)} - \frac{\Re k_{sp}}{|k_{sp}|^2 \rho}(|J_1^{(1)}|^2 + |Y_1^{(1)}|^2)]\}$$

The permittivity of silver at $\lambda=357$ nm is $\varepsilon_r=-2.145$ and $\varepsilon_i = 0.275$[32] where $\varepsilon_r$ and $\varepsilon_i$ are the real and imaginary parts of the permittivity, respectively. Given that $k_{sp} = k_0\sqrt{(\varepsilon_1\varepsilon_2)/(\varepsilon_1 + \varepsilon_2)}$ and $k_{1z} = \sqrt{k_0^2 - k_{sp}^2}$, one obtains $\Re k_{sp} = 1.353 k_0$, $\Im k_{sp} = .07 k_0$ and $k_{z1} = i0.914 k_0$, also we have $\Im k_{sp} \ll |k_{sp}|$. Substituting these values in (2-5) one can simplify it further to:

$$C = \frac{|\zeta|^2 \sin 2\varphi}{2\rho\omega}\{0.305[Y_1^{(1)} J_0^{(1)} - Y_0^{(1)} J_1^{(1)}] - 1.599[J_1^{(1)} J_0^{(1)} + Y_1^{(1)} Y_0^{(1)} - \frac{0.712}{\rho k_0}(|J_1^{(1)}|^2 + |Y_1^{(1)}|^2)]\} \quad (2\text{-}6)$$

**Supporting Note 3.**

The magnetic field of a circularly polarized dipole is

$$\boldsymbol{B} = \zeta e^{ik_{z1}z} e^{-i\omega t} e^{i\varphi} \{-i[\frac{k_{sp}}{k_{z1}} - \frac{1}{k_{sp}\rho}]\frac{H_1^{(1)}(k_{sp}\rho)}{\rho\omega}\hat{\rho} +$$

$$\frac{1}{\omega}[(H_0^{(1)}(k_{sp}\rho) - \frac{1}{k_{sp}\rho}H_1^{(1)}(k_{sp}\rho))k_{z1} + \frac{k_{sp}^2}{k_{z1}}\frac{\partial H_1^{(1)}(k_{sp}\rho)}{\partial(k_{sp}\rho)}]\hat{\varphi} \qquad (3\text{-}1)$$

$$+\frac{2}{\rho\omega}[(H_0^{(1)}(k_{sp}\rho) - \frac{1}{k_{sp}\rho}H_1^{(1)}(k_{sp}\rho))]\hat{z}\}$$

It is clearly seen that the magnetic field possesses longitudinal components as well as transverse components. The optical chirality is the equation 8 in the main text. Making use of the definition of the Hankel function and the expanded form of the wave-vectors one can rewrite the optical chirality as

$$C = \frac{|\zeta|^2}{2\rho\omega}\{\frac{-(J_1^2+Y_1^2)}{|k_{sp}|^2 \rho^2} + \frac{\Im k_{sp}(J_1J_0+Y_1Y_0)}{|k_{z1}|} + \frac{\Re k_{sp}(J_1J_0+Y_1Y_0)}{|k_{sp}|^2 \rho} - \frac{\Re k_{sp}(J_1Y_0-Y_1J_0)}{|k_{z1}|}$$

$$-\frac{\Im k_{sp}(J_1Y_0-Y_1J_0)}{|k_{sp}|^2 \rho} - \frac{\Im k_{sp}\Im k_{z1}(J_1J_0+Y_1Y_0)}{|k_{sp}|^2} - \frac{\Re k_{sp}\Im k_{z1}(J_1Y_0-Y_1J_0)}{|k_{sp}|^2} \qquad (3\text{-}2)$$

$$+\frac{(J_1J_0+Y_1Y_0)}{|k_{z1}|^2|k_{sp}|^2}[3(\Re k_{sp})^2\Im k_{sp}\Im k_{z1} - (\Im k_{sp})^3\Im k_{z1}] + \frac{(J_1Y_0-Y_1J_0)}{|k_{z1}|^2|k_{sp}|^2}[3(\Im k_{sp})^2\Re k_{sp}\Im k_{z1} + (\Re k_{sp})^3\Im k_{z1}]$$

Despite the complexity of equation (3-2) one notices that this equation can be simplified to a great extent depending on the working wavelength, as both real and imaginary parts of the $k_{sp}$ are explicitly frequency dependent. Applying the derived optical constants of the silver at λ=357 nm as was done in Supporting Note 2 we observe that $\Im k_{sp} \ll \Re k_{sp}, \Im k_{sp} \ll \Im k_{z1} \approx |k_{sp}|$ and equation (3-2) reduces to

$$C = \frac{|\zeta|^2}{2\rho\omega}\{\frac{-(J_1^2+Y_1^2)}{|k_{sp}|^2 \rho^2} + [\frac{3(\Re k_{sp})^2\Im k_{sp}\Im k_{z1}}{|k_{z1}|^2|k_{sp}|^2} + \frac{\Re k_{sp}}{|k_{sp}|^2 \rho}](J_1J_0+Y_1Y_0)$$

$$+[\frac{(\Re k_{sp})^3\Im k_{z1}}{|k_{z1}|^2|k_{sp}|^2} - \frac{\Re k_{sp}\Im k_{z1}}{|k_{sp}|^2} - \frac{\Re k_{sp}}{|k_{z1}|}](J_1Y_0-Y_1J_0)\} \qquad (3\text{-}3)$$

**Supporting Note 4.**

Calculating the optical chirality of a set of electric-magnetic dipoles in free space is analytically feasible. For a combination of arbitrarily oriented electric and magnetic dipoles one can obtain the electric and magnetic fields in free space to be a superposition of the electromagnetic fields of each dipole. More explicitly the electric and magnetic fields of a general electric dipole, $\vec{p}$, are[48]:

$$E(r) = \frac{1}{4\pi\varepsilon_0}\{k^2\frac{e^{ikr}}{r}(\vec{n}\times\vec{p})\times\vec{n} + [3\vec{n}(\vec{n}.\vec{p})-\vec{p}](\frac{1}{r^3}-\frac{ik}{r^2})e^{ikr}\}$$

$$H(r) = \frac{ck^2}{4\pi}(\vec{n}\times\vec{p})\frac{e^{ikr}}{r}(1-\frac{1}{ikr})$$

and those of a magnetic dipole, $\vec{m}$, are:

$$E(r) = \frac{-z_0}{4\pi}k^2(\vec{n}\times\vec{m})\frac{e^{ikr}}{r}(1-\frac{1}{ikr})$$

$$H(r) = \frac{1}{4\pi\varepsilon_0}\{k^2\frac{e^{ikr}}{r}(\vec{n}\times\vec{m})\times\vec{n} + [3\vec{n}(\vec{n}.\vec{m})-\vec{m}](\frac{1}{r^3}-\frac{ik}{r^2})e^{ikr}\}$$

here $\vec{n}$ is the unit vector in the direction of the line that connects origin and the observation point. Thus for any combination of such dipole one will have:

$$E(r) = \frac{-z_0}{4\pi}k^2(\vec{n}\times\vec{m})\frac{e^{ikr}}{r}(1-\frac{1}{ikr}) + \frac{1}{4\pi\varepsilon_0}\{k^2\frac{e^{ikr}}{r}(\vec{n}\times\vec{p})\times\vec{n} + [3\vec{n}(\vec{n}.\vec{p})-\vec{p}](\frac{1}{r^3}-\frac{ik}{r^2})e^{ikr}\}$$

$$H(r) = \frac{ck^2}{4\pi}(\vec{n}\times\vec{p})\frac{e^{ikr}}{r}(1-\frac{1}{ikr}) + \frac{1}{4\pi\varepsilon_0}\{k^2\frac{e^{ikr}}{r}(\vec{n}\times\vec{m})\times\vec{n} + [3\vec{n}(\vec{n}.\vec{m})-\vec{m}](\frac{1}{r^3}-\frac{ik}{r^2})e^{ikr}\}$$

If we calculate optical chirality for such a system, after some algebra, we obtain:

$$C = \frac{\omega}{8\pi^2c^2\varepsilon_0}\text{Im}\{\frac{k^4}{r^2}[\vec{p}.\vec{m}-(\vec{n}.\vec{p})(\vec{n}.\vec{m})] - \frac{2k^2}{r^4}(\vec{p}.\vec{m}) + \frac{2k^2}{r^4}(\vec{n}.\vec{p})(\vec{n}.\vec{m}) + (\frac{1}{r^6}+\frac{k^2}{r^4})[\vec{p}.\vec{m}+3(\vec{n}.\vec{p})(\vec{n}.\vec{m})]\}$$

**Supporting Figure 1.**

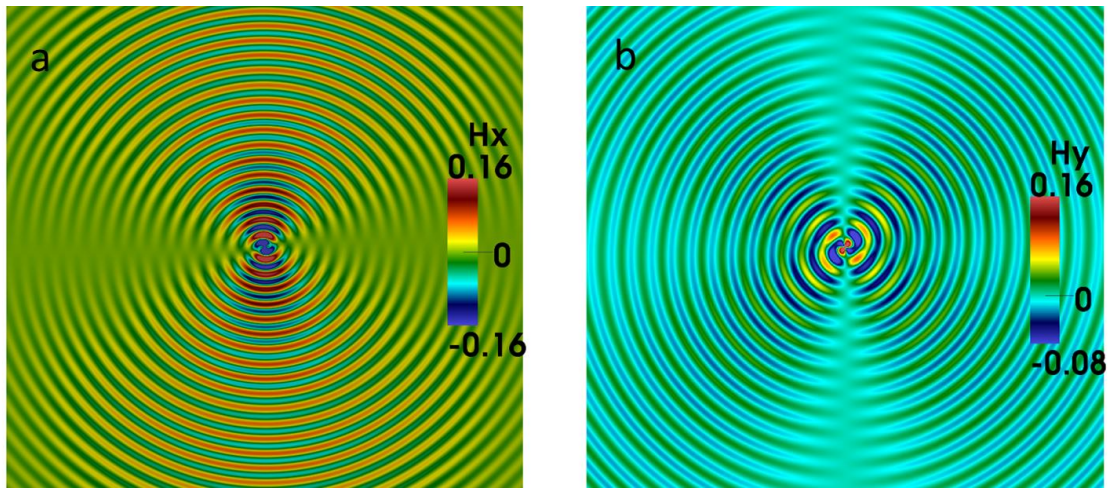

**Figure S1.** (a) x and (b)y components of the magnetic field for a chiral emitter at λ=357 nm. It is seen that the two components yield a non-zero field everywhere on the plane.